\documentstyle[twocolumn,prl,aps]{revtex}
\large
\begin{document}
\draft
\twocolumn[
\hsize\textwidth\columnwidth\hsize\csname @twocolumnfalse\endcsname

\title
      {
       Intra-molecular origin of the fast relaxations observed in \\the 
       Brillouin light scattering spectra of molecular glass-formers
      }
\author{
        G.~Monaco$^{1}$,
        S.~Caponi$^{2}$,
        R.~Di~Leonardo$^{3}$,
        D.~Fioretto$^{2}$,
        G.~Ruocco$^{3}$,
       }
\address{
         $^1$
         European Synchrotron Radiation Facility, B.P. 220, F-38043 
         Grenoble Cedex, France. \\
         $^2$
         INFM and Dipartimento di Fisica, Universit\`a di Perugia,
	 I-06100, Perugia, Italy. \\
         $^3$
         INFM and Dipartimento di Fisica, Universit\'a di L'Aquila,
	 I-67100, L'Aquila, Italy. \\
        }

\date{\today}
\maketitle
\begin{abstract}

The Brillouin light scattering spectra of the $o$-terphenyl single 
crystal are compared with those of the liquid and the glass phases. 
This shows: 
i) the direct evidence of a fast relaxation at frequencies
$\nu$$\approx$5 GHz in both the single crystal and the glass; 
ii) a similar temperature dependence for the attenuation 
of the longitudinal sound waves in the single crystal and the glass; and 
iii) the absence of coupling between the fast relaxation and the 
transverse acoustic waves.
These results allow us to assign such a relaxation to the coupling 
between the longitudinal acoustic waves and intra-molecular vibrations, 
and therefore to exclude any relationship between it and the glass 
transition.  
\end{abstract}


]
The study of the atomic dynamics of systems undergoing the
structural liquid to glass transition is nowadays a very
fertile topic in the condensed matter physics, and is stimulating
many relevant developments on both the experimental and the
theoretical side \cite{review}. 
This study is historically characterized by the attempt to extend 
our knowledge of the dynamics to larger and larger time ranges. 
Traditionally, this attempt has been focused on the time range 
of the structural relaxation, which shows prominent spectral changes 
in the glass transition region. In more recent years, conversely, 
much of the attention has shifted towards shorter times, i.e. 
towards the so-called fast processes range. 

The interest in extending as much as possible the probed time window 
has been much stimulated by the development of the Mode Coupling 
Theory (MCT) \cite{MCT}, which describes the time evolution of the 
$q$-component of the density-density correlation function, $F(q,t)$, 
of a {\it simple} liquid when its temperature is changed across the 
metastable undercooled region. In fact, the MCT makes, among the others, 
specific predictions on the time range comprising the earliest part of 
the structural relaxation and the latest part of the fast processes,
the $\beta$-region \cite{MCT}. 

The MCT predictions have been the subject of extensive experimental 
studies \cite{Pisa}. Among them, the ones which really probe 
$F(q,t)$ (or its time Fourier transform, the dynamic structure 
factor, $S(q,\omega)$) are based on inelastic neutron scattering 
(INS) in the 10 nm$^{-1}$ $q$ range \cite{neutrons}, and on 
Brillouin light scattering (BLS) in the 10$^{-2}$ 
nm$^{-1}$ $q$ range \cite{light}. In these studies, the relevance of 
the short-times region for a proper spectral description has been 
underlined; this region has been associated to the MCT $\beta$-region, 
and several MCT predictions have been tested, e.g. the cusp-like 
discontinuity in the Debye-Waller factor at the dynamical transition 
temperature, $T_c$ \cite{neutrons,light}.
Unfortunately, while the MCT predictions 
are devoloped for ideal systems (hard spheres or monoatomic systems), 
the glass formers which are the object of experimental 
studies are mostly molecular compounds. These latter ones are 
obvously characterized by an intra-molecular dynamics which, in 
principle, could just be effective in the $\beta$-region \cite{Herzfeld}.

Some of us have recently reported on a fast relaxational dynamics in the 
10$^{-11}$ s time range which appears in the $S(q,\omega)$ spectra 
obtained by BLS in glassy OTP \cite{fast}. This
fast dynamics has also been analyzed using a phenomenological ansatz 
for the memory function and has been found to be active in both 
the glass and the liquid, and to be essentially temperature 
independent \cite{giulio}. Relaxation processes in a similar time window
had been observed in previous studies on similar glass-formers \cite{light}, 
and had been described in terms of the MCT $\beta$-region. 

In order to clarify the nature of this relaxation, we have performed 
on an OTP single crystal a further BLS experiment, on the results of which 
we report in this paper. As a matter of fact, we observe in the spectra
of the single crystal the same fast relaxation which was observed 
in both the liquid and the glass \cite{giulio,fast}. As a consequence,
we conclude that this fast relaxation is {\it not} related to the glass 
transition and should not be associated with the MCT $\beta$-region.
Moreover, by comparing the longitudinal (L) and transverse (T) sound 
attenuation in the liquid and the glass with the L sound attenuation in 
the single crystal, we also conclude that the fast relaxation arises from 
the coupling of density fluctuations to intra-molecular degrees of freedom. 

The OTP single crystal has been prepared from $99\%$ purity powder 
(Aldrich Chemicals) by ricrystallization in a methanol solution. The 
typical dimensions of the 
obtained needles-shaped crystals are 2x2x10 mm$^3$. The crystals have been 
characterized by X-ray diffractometry, and the measured lattice parameters 
of the orthorhombic unit cell (P$2_1 2_1 2_1$ symmetry) are in agreement 
with previous determinations 
($a=6.024$ $\AA$, $b=11.729$ $\AA$, and $c=18.582$ $\AA$) 
\cite{structure}.  The main axis of the needles is along the $a$ axis, 
and the lateral faces are found to be parallel to the 011 and the 023 
lattice planes. 

The BLS spectra of the single crystal have been collected in backscattering 
geometry, with the exchanged wavevector $q$ along the (001) direction. 
In the chosen configuration, only the L Brillouin peaks are allowed.
As the exciting line, we have used the single mode from a 
Coherent compass 532/400 laser operating 
at $\lambda_o$=532 nm with a typical power of $\approx$300 mW. 
The scattered light has been analyzed by a Sandercock-type (3+3)-pass tandem 
Fabry-Perot frequency analyzer, with a finesse of $\approx$100 and a contrast 
$>$5$\times $10$^{10}$ \cite{tandem}, set to a free spectral range of 
$\approx$10 GHz. No analysis in polarization has been done, and the 
measurements have been performed in the 37$\div$303 K temperature range (the 
melting temperature is $T_m$=329 K).
The same experimental setup has been used to measure the Brillouin peaks in 
the OTP glass in the temperature range 27$\div$150 K (the glass transition
temperature is $T_g$=244 K). 
Moreover, further measurements in the 90$^o$ scattering geometry and with a VH 
polarization configuration have been collected in the 250$\div$310 K 
temperature range in order to reveal the T Brillouin peaks.
A dilute aqueous suspension of Latex particles (120 nm in diameter) has been 
used to determine the instrumental resolution function. In the 90$^o$ geometry 
measurements, the linewidths of the T Brillouin peaks are affected not only by 
the resolution function of the 
spectrometer, but also by the finite solid angle 
which is defined by the collection optics. This effect, negligible for 
scattering angles close to 180$^o$, 
gives a non vanishing contribution at 90$^o$, 
and must be carefully taken into 
account in order to extract the true linewidth. 
The deconvolution procedure used for this purpose is described 
elsewhere \cite{silvia}.

The room temperature ($T$=296 K) spectrum of the OTP single crystal with $q$
along the (001) direction is presented on a log-linear scale in the lower part 
of Fig.~1. Apart from the small leakage 
of the quasi-transverse modes, observed 
at $\approx$10 GHz and due to the residual misalignement from the (001) 
direction, this spectrum shows 
clear signatures of anharmonic effects: 
broad Brillouin peaks and a broad central band. In the upper part of the same 
figure, the spectrum of the supercooled liquid at $T$=245 K is also reported, 
for comparison. The two spectra have been collected under identical 
experimental conditions and, at the chosen temperatures, show 
comparable linewidths for the Brillouin peaks. The
similarity between the two spectra is remarkable, in particular for what 
concerns the central region which, in the supercooled liquid and in the glass, 
has been described in terms of a secondary relaxation \cite{giulio,fast}.
This figure by itself already suggests that this relaxation arises
from the anharmonic coupling between the acoustic waves and other degrees of
freedom.

In order to have a deeper insight into this anharmonic process, we have 
measured the temperature dependence of the single crystal BLS spectra. 
In Fig.~2 we present few spectra at 
selected temperatures. Here, in order to emphasize the temperature dependence 
of the linewidth, only the Brillouin peaks are reported. On cooling, 
a strong 
reduction of the linewidth can directly be observed.

A quantitative description of the reduction of the Brillouin linewidth with 
temperature has been obtained by fitting the 
peaks region of the current spectra 
(i.~e. of the measured spectra multiplied by the frequency squared) 
with a Lorentzian lineshape convoluted to 
the experimental resolution function. 
The values of $2 \Gamma_c$ (FWHM) resulting 
from the fits are reported in Fig.~3 
($\diamond$) in the 55-303 K temperature range. 
At temperatures lower than 55 K the Brillouin linewidth is almost 
indistinguishible from that of the 
instrumental resolution function, and thus 
$\Gamma_c$ could not be reliably obtained. In the same figure the similar 
quantities, $2 \Gamma_d$, measured in the 
liquid and glassy OTP are also reported 
($\bullet$): the data above 114 K are taken from 
Ref.~\cite{giulio}, and we have 
here extended in temperature those measurements 
down to 27 K.  The inset reports 
the same data on a larger $T$-range. 

The comparison between the Brillouin linewidth of the single crystal and the
liquid/glass comes out to be very instructive. 
At temperatures above $T_g$, in the undercooled and normal liquid phase, the 
linewidth is dominated by the structural relaxation which, however, is 
expected to be of no influence on $\Gamma_d$ below $T_g$. Here, however, 
$\Gamma_d$ is still temperature dependent, and this has been attributed to 
the presence of a secondary, fast relaxation \cite{giulio,fast}. On cooling, 
eventually, $\Gamma_d$ seems to approach a limiting value, $\Gamma_o$. This 
latter value is here estimated to be $2 \Gamma_o$=70$\pm$10 MHz by a quadratic 
fit on the data below 150 K. The present value is slightly smaller than that 
reported in Ref.~\cite{giulio,fast} (120$\pm$20 MHz), where it was inferred by 
using the $\Gamma_d$ values for $T >$114 K 
and by scaling high frequency Inelastic 
X-ray Scattering (IXS) data \cite{otpx}. 
From the comparative analysis of the IXS 
and the BLS data, $\Gamma_o$ has been attributed 
to the presence of topological 
disorder in the glass \cite{giulio,fast}. 
To sum up, the linewidth of the liquid 
and glassy OTP has been described as the sum of three different contributions: 
$\Gamma_d (T)=\Gamma_o+\Gamma_f(T)+\Gamma_\alpha(T)$, where 
$\Gamma_\alpha(T)$ is the contribution 
of the structural relaxation and becomes 
negligible below $T_g$, $\Gamma_f(T)$ is 
the contribution of the secondary, fast 
relaxation, and $\Gamma_o$ is the contribution of the topological disorder.
Within this picture, we can get a direct 
estimate of $\Gamma_f (T)$ below $T_g$ by 
simply subtracting $\Gamma_o$ from the total linewidth $\Gamma_d$. 
The difference data are also reported in Fig.~3 ($\oplus$), and they represent 
the only contribution of the secondary relaxation to the total linewidth.

The present measurements on the OTP single 
crystal not only strongly support the
previous picture, but furthermore 
give us a fundamental hint to understand the 
nature of the secondary relaxation. In fact, as 
reported in Fig.~3, the linewidth 
of the single crystal, $\Gamma_c$, shows a $T$-dependence which, below $T_g$,  
closely follows that of $\Gamma_d$ and, 
furthermore, nicely agrees with that of
$\Gamma_f$. Thus, we can conclude that: i) the description of
the linewidth as the sum of the previously 
discussed contributions is a reasonable 
and physical ansatz, and ii) the anharmonic process observed in the single 
crystal and the fast process observed in the 
liquid and glass \cite{giulio,fast} do 
share a common origin. 
As a consequence, we rule out the possibility 
that the secondary process observed 
in OTP, both in the glass and in the liquid, is related to the $\beta$-region 
discussed by the MCT, and, more generally, 
that this process is related to the 
glass transition itself.

Left open is the question of the origin 
of the fast process. In particular one
can associate this process to the coupling of the longitudinal acoustic waves
either with i) intra-molecular degrees of freedom (vibrational relaxation) 
or with ii) other inter-molecular modes. 
A key experiment to solve this ambiguity 
is the determination of the absorption 
of the transverse sound waves. Indeed, in case ii) one expects a 
transverse Brillouin linewidth comparable 
to the longitudinal one around $T_g$, 
and in any case with a similar $T$ dependence. 
On the contrary, the anharmonicity
of type i) is expected to give a negligible contribution to the linewidth 
as the vibrational relaxations strongly 
couple only to the density fluctuations 
\cite{zwanzig}, and the coupling with the shear waves can take place only 
through the much less efficient mechanism of the translation-rotation 
coupling \cite{Berne}.

The comparison of the the L and T sound absorption
is presented in Fig.~4, where we report the linewidths normalized by $q^2$,
$D=4\pi\Gamma/q^2$ 
(apparent kinematic viscosities, $D_L$ and $D_T$ for the L and T Brillouin
peaks, respectively) in order to take into account the trivial 
contribution due to the different scattering geometries used to measure the 
L ($\bullet$) and the T ($\triangle$) Brillouin peaks 
We also report the results of previous literature mesurements for
the T linewidths ($\star$) \cite{steffen}; the agreement between the whole 
set of $D_T$ data can be observed to be good. 
It is immediately seen from Fig.~4 that $D_T$,
dominated by the contribution of the $\alpha$ process above $T_g$,
does not follow $D_L$, and, in particular, that, on cooling, it drops to
vanishingly small values at $T_g$. This is a 
clear indication that the transverse 
waves are not affected by the fast process and, 
therefore, that this process must 
be classified as vibrational in nature. 

It has to be emphasised that the presence 
of such a kind of secondary processes
in a molecular system like OTP is not surprising. In fact, the class of the 
fragile glass-formers (as OTP \cite{Angell}) 
has a strong overlap \cite{giulio} 
with the so-called Kneser liquids \cite{Herzfeld}. These systems have been 
widely studied by ultrasonics, and they are known to be 
characterized by the presence of secondary relaxations in the 
10$^{-9}$ $\div$ 10$^{-11}$ s time range.
Moreover, the presence of similar relaxation 
processes has also been observed in 
BLS spectra of analogous systems, and in particular 
those characterized by the 
presence of phenyl rings \cite{fytas}. 

In conclusion, by comparing the BLS spectra of the single crystal and the 
liquid/glass OTP, we have shown that the secondary, fast relaxation process 
previously observed in the OTP glass and liquid \cite{giulio,fast} is active 
also in the single crystal. This observation rules out the possibility that 
this process is related to the glass transition. Moreover, by comparing the 
longitudinal and transverse acoustic attenuation we have shown that the fast 
process does not couple to the tranverse sound waves, and thus we attribute 
it to a vibrational relaxation of intra-molecular degrees of freedom. 

Given this picture, it is natural to wonder 
whether the presence of a vibrational 
relaxation in the 10$^{-11}$ s time range 
is a peculiarity of OTP. Actually, the 
large ultrasonic literature on related 
systems \cite{Herzfeld} would suggest that 
this kind of relaxations couple to the density fluctuations in a wide class 
of molecular liquids, in particular the 
fragile glass-formers. The presence of 
vibrational relaxations in real molecular 
systems could completely hinder features 
intrinsic to the glass transition as, 
for example, the MCT $\beta$-region. Thus, it 
could come out to be dangerous, whether not incorrect, to describe the fast 
relaxation region of the memory function used 
to describe BLS spectra in terms of 
a simple MCT approach, as it has often been 
done \cite{light}. This would certainly 
be the case for OTP.

We thank A.~Phany and P.~F.~Zanazzi for their 
help in the crystal orientation.

\begin{center}
\footnotesize{\bf FIGURE CAPTIONS} 
\end{center}

{
\footnotesize {

\begin{description}
\item   {FIG. 1 - 
BLS spectra in backscattering geometry of the single crystal 
($q$ along the (001) direction) and of the strongly supercooled OTP at 
the indicated temperatures. 
}  

\item {FIG. 2 -
BLS spectra of the OTP single crystal at the following temperatures 
(from high to low peak position): 37, 96, 138, 194, 251, 303 K.
The spectra have been collected in the backscattering geometry with 
the exchanged $q$-vector along the (001) direction. The overall 
instrumental resolution was 100 MHz FWHM.
} 

\item {FIG. 3 -
Temperature dependence of the measured linewidths in the OTP single crystal 
($\diamond$) and glass/liquid ($\bullet$). The inset shows the same data on a 
larger temperature range \cite{giulio}. The symbol $\oplus$ represents the 
only contribution of the secondary, fast relaxation to the linewidth in the 
glass.
}

\item {FIG. 4 -
Temperature dependence of Brillouin linewidths normalized by $q^2$: 
longitudinal data from Ref.~\cite{giulio}) ($\bullet$); T data from this
work ($\triangle$), and literature data from Ref.~\cite{steffen} ($\star$).
The dashed line indicates the low temperature limiting value of the 
longitudinal apparent kinematic viscosity, $D_{o}=4\pi\Gamma_{o}/q^2$.
}

\end{description}
}
}
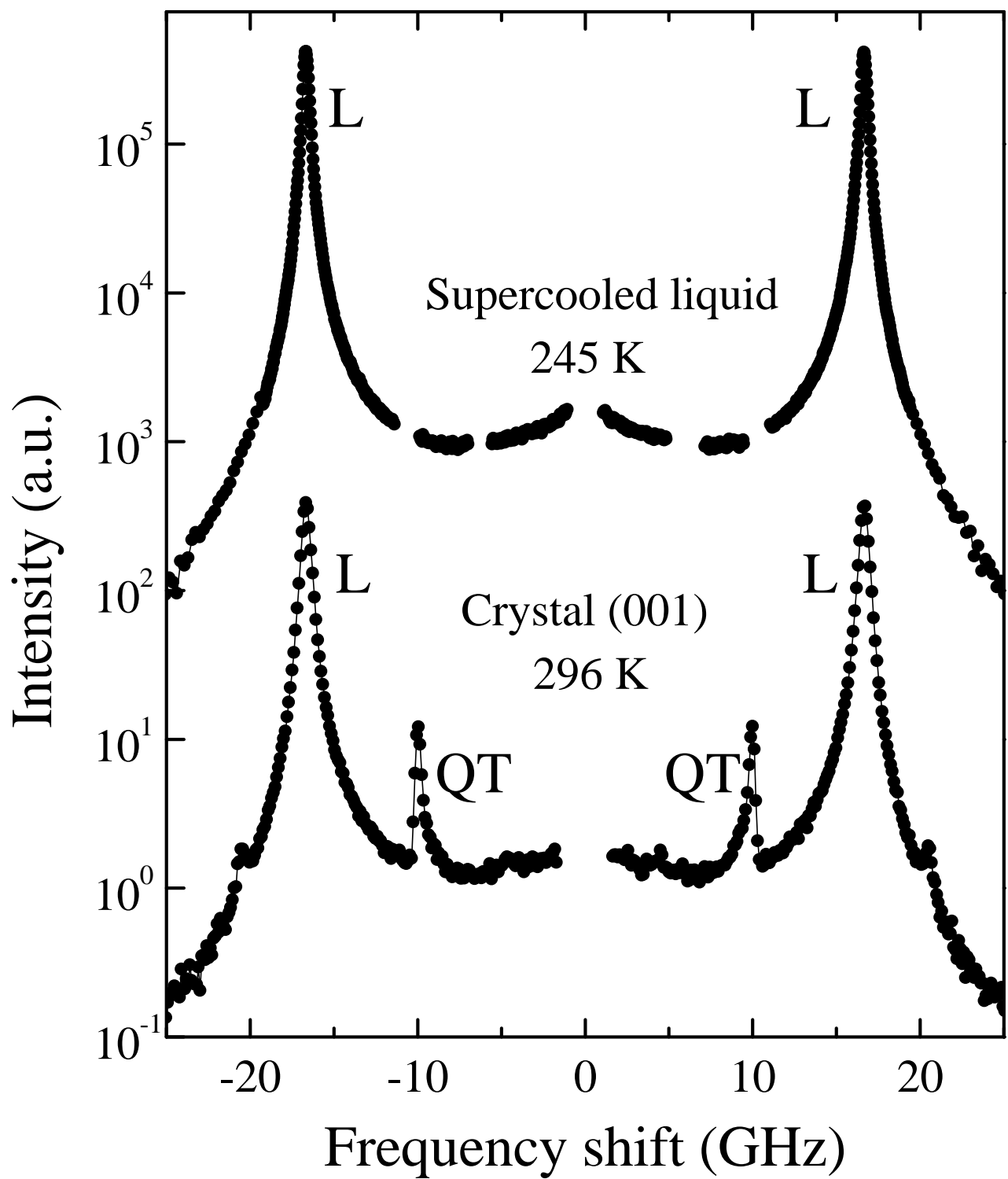
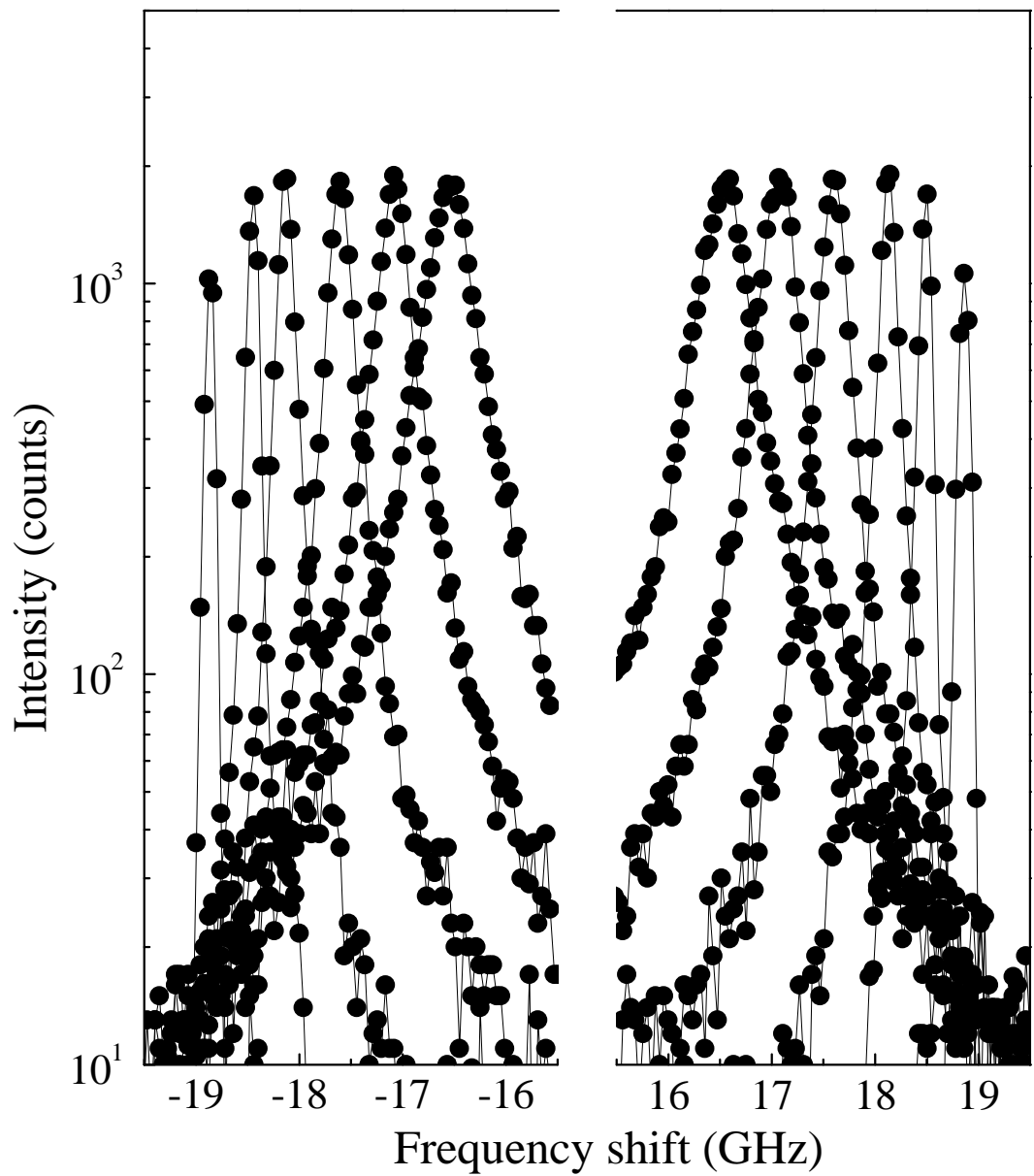
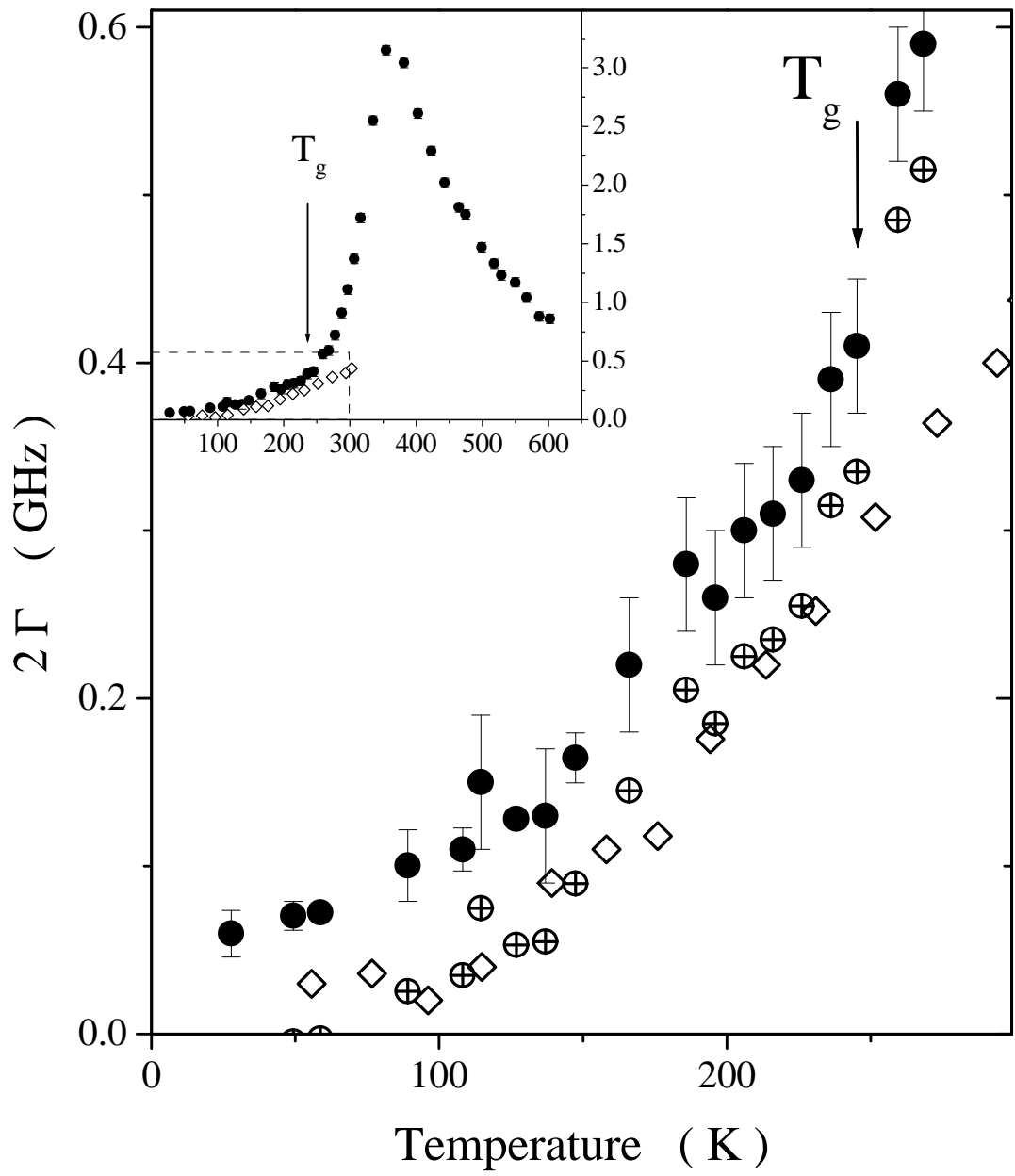
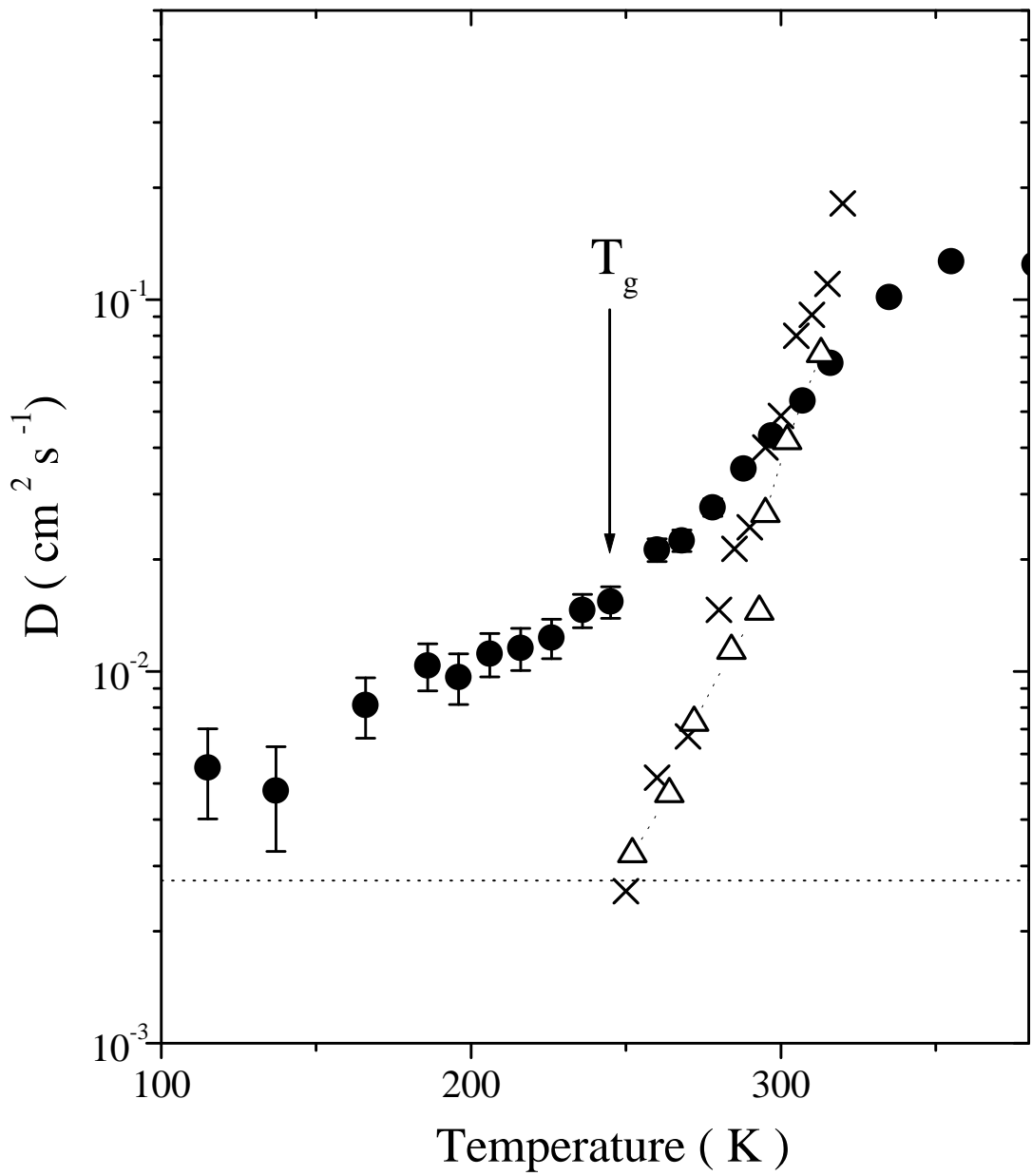
\end{document}